\documentclass[aps,pra,reprint,superscriptaddress]{revtex4-2}

\usepackage[english]{babel}
\usepackage{graphics}
\usepackage{array}
\usepackage{graphicx}
\usepackage{todonotes}
\usepackage{amsfonts,amssymb,amsmath}
\usepackage{color}
\usepackage{threeparttable}
\usepackage{physics}
\usepackage{comment}
\usepackage{orcidlink}

\usepackage[separate-uncertainty,per-mode=fraction]{siunitx}%,exponent-product=\cdot
\usepackage[extra]{optional} % choose if extra text or noextra.

\definecolor{darkblue}{rgb}{0.0,0.0,0.4}
\definecolor{darkgreen}{rgb}{0.0,0.4,0.0}
\definecolor{darkred}{rgb}{0.6,0.0,0.0}
\hypersetup{colorlinks,linkcolor=darkblue,citecolor=darkblue,urlcolor=darkblue}

\newcolumntype{M}[1]{>{\centering\arraybackslash}m{#1}}
\newcommand\cedilla{}
\let\cedilla\c
\DeclareUnicodeCharacter{0327}{\c{a}}

\begin{document}
\title{Decoupled sound and amplitude modes in trapped dipolar supersolids}

\author{J. Hertkorn \orcidlink{0000-0002-5377-9981}}
\email{jhertkorn@pi5.physik.uni-stuttgart.de}
\affiliation{5. Physikalisches  Institut  and  Center  for  Integrated  Quantum  Science  and  Technology,Universit\"at  Stuttgart,  Pfaffenwaldring  57,  70569  Stuttgart,  Germany}

\author{P. Stürmer \orcidlink{0000-0002-1447-8545}}
\affiliation{Division of Mathematical Physics and NanoLund, LTH, Lund University, Box 118, SE-221 00 Lund, Sweden }
\author{K. Mukherjee \orcidlink{0000-0002-1182-2785}}
\affiliation{Division of Mathematical Physics and NanoLund, LTH, Lund University, Box 118, SE-221 00 Lund, Sweden }

\author{K.S.H. Ng}
\author{P. Uerlings}
\author{F. Hellstern}
\author{L. Lavoine}
\affiliation{5. Physikalisches  Institut  and  Center  for  Integrated  Quantum  Science  and  Technology,Universit\"at  Stuttgart,  Pfaffenwaldring  57,  70569  Stuttgart,  Germany}

\author{S.M. Reimann \orcidlink{0000-0003-1869-9799}}
\affiliation{Division of Mathematical Physics and NanoLund, LTH, Lund University, Box 118, SE-221 00 Lund, Sweden }
\author{T. Pfau}

\author{R. Klemt \orcidlink{0009-0002-3231-447X}}
\email{rklemt@pi5.physik.uni-stuttgart.de}
\affiliation{5. Physikalisches  Institut  and  Center  for  Integrated  Quantum  Science  and  Technology,Universit\"at  Stuttgart,  Pfaffenwaldring  57,  70569  Stuttgart,  Germany}

\begin{abstract}
We theoretically investigate elementary excitations of dipolar quantum gases across the superfluid to supersolid phase transition in a toroidal trap. We show how decoupled first sound, second sound, and Higgs modes emerge by following their origin from superfluid modes across the transition. The structure of these excitations reveals the interplay between crystal and superfluid oscillations. Our results unify previous notions of coupled Goldstone and Higgs modes in harmonic traps, allowing us to establish a correspondence between excitations of trapped and infinitely extended supersolids. We propose protocols for selectively probing these sound and amplitude modes, accessible to state-of-the-art experiments.
\end{abstract}

\maketitle

Spontaneous symmetry breaking permeates our understanding of nature and is central to concepts ranging from the emergence of elementary particles to phase transitions in condensed matter~\cite{Anderson1972}. These phase transitions are often signaled by a characteristic spectrum of Goldstone and Higgs collective excitation modes across criticality. In superfluids at zero temperature, the Goldstone mode emerges as superfluid sound from breaking the $U(1)$ symmetry 
\cite{PitaevskiiBook2016,Schmitt2015Superfluidity}. The coupling to another (non-superfluid) component can then lead to a second sound mode. This concept historically originated from a hydrodynamic two-fluid model describing the 
out-of-phase oscillation between thermal and superfluid components of Helium~\cite{Atkins1959,Donnelly2009,Yan2024}. 
In a supersolid~\cite{Gross1957, Yang1962, Boninsegni2012}, the coexistence of superfluidity and density modulation results in a second Goldstone mode at zero temperature and a novel type of second sound. Corresponding theoretical investigations of sound in Helium have a long history \cite{Balibar2007, *Prokofev2007,  *Balibar_2008}, yet in experiments to date the supersolid phase of helium remains elusive  \cite{Kim2004a, *Kim2004b, *Balibar2010, *Chan2013}. Supersolids were first realized in the context of ultracold gases, using BECs with spin-orbit coupling \cite{Li2017,Hirthe2024,Geier2023} or in multimode optical cavities \cite{Leonard2017}, as well as dipolar quantum gases \cite{Tanzi2019, Fabian2019, Chomaz2019}. In the latter system, indications of a Goldstone mode associated with translational symmetry breaking were observed~\cite{Guo2019,Stringari2019}. Despite considerable efforts, the identification of second sound, its relationship with the Goldstone mode and the development of a consistent picture describing the different sound modes has proven challenging, both experimentally and theoretically  \cite{Hertkorn2019,Guo2019,Tanzi2019,Hertkorn2021Fluctuation,Hertkorn2021Spectrum2D,Tanzi2021,Natale2019,Mukherjee2023LinChain}. In addition, hydrodynamic theories, which can identify the two sound modes, are unable to capture gapped modes such as the Higgs mode~\cite{Dorsey2010, Hofmann2021, Stringari2023Ring, Platt2024}, which is yet to be observed experimentally in a supersolid. The search is typically complicated by harmonic trap confinement, where elementary excitations couple to center-of-mass (COM) modes. Here,  toroidal potentials \cite{Amico2021,Tengstrand2021,Bland2022,Bland2020,Tengstrand2023} are a better choice: The continuous rotational symmetry and periodic boundary conditions along the azimuthal direction link experimentally achievable finite-size supersolids with those in the bulk addressed by mean-field  
\cite{Ferlaino2023Elongated, Platt2024} and quantum Monte Carlo methods \cite{Saccani2012,Cinti2014}.

In this work, we show how first sound, second sound, and Higgs amplitude modes emerge in dipolar supersolids by calculating the elementary excitation spectrum in a toroidal trap across criticality. We classify the spatial symmetries of the eigenmodes using group theory and show how a quasimomentum $\tilde{q}$ is assigned to them. This establishes a direct correspondence between the excitations of finite-sized systems and infinite systems in which the description using band structures becomes meaningful. We show that the second (first) sound is an out-of-phase (in-phase) oscillation between superfluid and crystal components and that both sound modes emerge from the superfluid sound branch near the roton momentum $q_\mathrm{rot}$ which defines the crystal periodicity. At the critical point, $q_\mathrm{rot}$ becomes the $\tilde{q}=0$ point of the emergent Brillouin zone, at which the superfluid sound mode splits into the $\tilde{q}=0$ instance of the second sound branch -- the zero-energy Goldstone mode -- and a separate isolated Higgs amplitude mode.   
The analysis of these modes allows us to design experimental protocols for probing single elementary excitations of the supersolid and more generally, to define selection rules of the supersolid excitation spectrum.

We numerically calculate the ground state (GS) of our atomic dipolar system using the extended Gross-Pitaevskii (eGPE) framework \cite{Ronen2006,Saito2016,Ferrier-Barbut2016,Wenzel2017,Roccuzzo2019}, which includes quantum fluctuations to leading order in a local density approximation \cite{Schutzhold2006,Lima2011,Lima2012,Petrov2015,Ferrier-Barbut2016}. The contact and dipolar two-body interaction strengths of the $N$ atom system are controlled by the s-wave scattering length $a_s$ and dipolar length $a_{dd} = \mu_0 \mu_d^2 M / 12 \pi \hbar^2$, respectively. Here $\mu_d$ denotes the magnetic moment and $M$ the atomic mass. We introduce a toroidal trap geometry by setting the external trapping potential to $V_t=M\omega_r^2\left[ (\rho-\rho_0)^2+\lambda^2 z^2\right]/2$, where $\rho_0$ denotes the radius of the torus, $\omega_r$ the radial trap frequency and $\lambda$ the aspect ratio between transversal and radial confinement \cite{Tengstrand2021}. We perform the 3D GS search using imaginary time evolution \cite{Bao2010} and conjugate gradient \cite{Modugno2003,Ronen2006,Antoine2017,Antoine2018} techniques. On top of the GS, we study the elementary excitations within the Bogoliubov-de Gennes (BdG) formalism described in detail e.\,g. in Refs.~\cite{Ronen2006,Hertkorn2019}. It is based on a linear expansion in the small amplitude parameter $\alpha$ of the wavefunction ${\psi(\boldsymbol{r},t) = \left\{ \psi_0(\boldsymbol{r}) + \alpha \left[ u_i(\boldsymbol{r}) e ^{-i\omega t} + v_i^{\ast}(\boldsymbol{r})e^{i\omega t} \right] \right\} e^{-i\mu t/\hbar}}$, with $\psi_0$ the GS wavefunction at a chemical potential $\mu$ and $i$ labeling the $i^\text{th}$ excitation. With this ansatz, we arrive at a system of linear BdG equations. Solving them numerically, we obtain the Bogoliubov amplitudes $u_i,\,v_i$ as well as the linear combination $f_i = u_i + v_i$ which relate to the density fluctuations of the GS density $n(t) = |\psi_0|^2 + 2\alpha \delta n_i \cos(\omega_i t)$, where $\delta n_i = \psi_0 f_i$. Due to the finite system size, the spectrum of elementary excitations is discrete.

\begin{figure}[tb!]
	\includegraphics[trim=0 0 0 0,clip,scale=0.55]{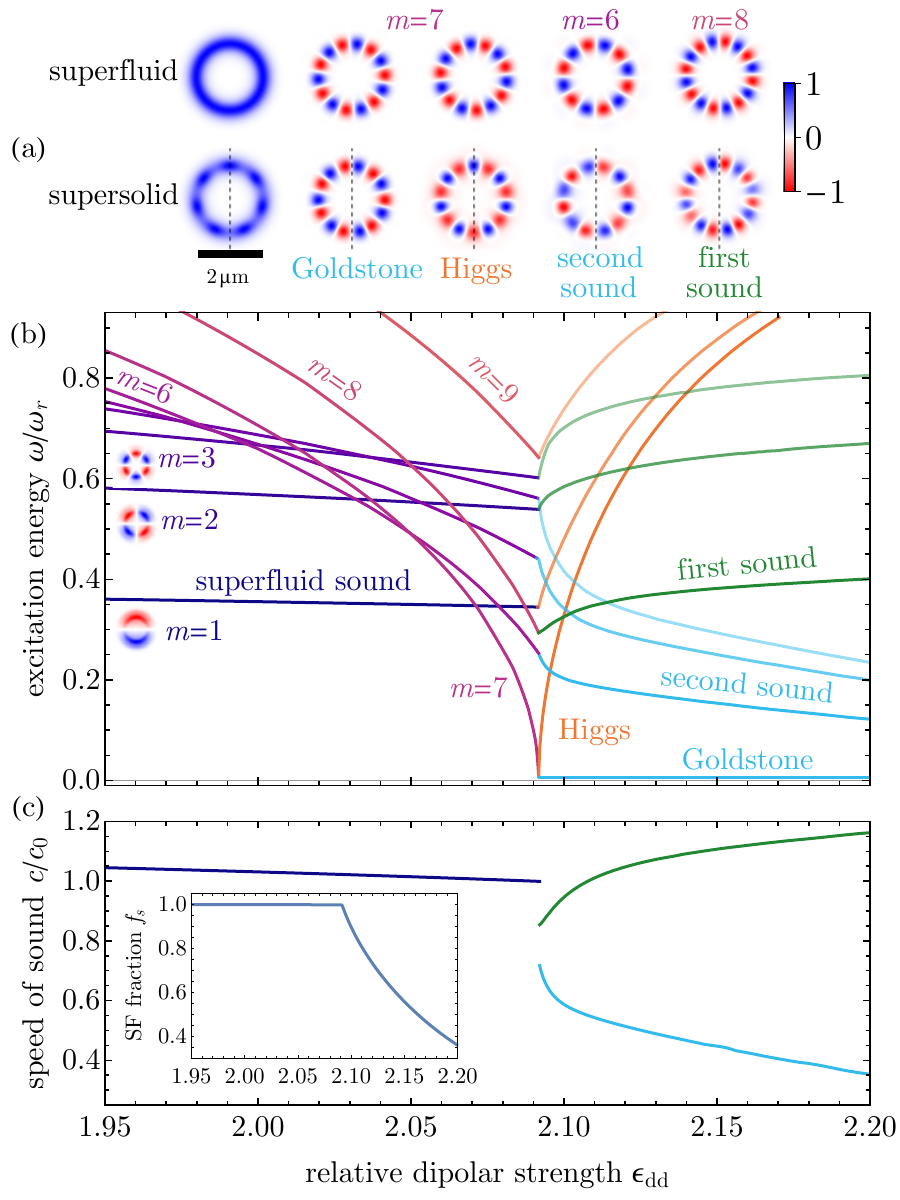}
	\caption{\textbf{Elementary excitations across the superfluid-to-supersolid phase transition.} (a) GS density ${n(x,y,z=0)}$ and Bogoliubov mode patterns ${f_i(x,y,z=0)}$ in arbitrary units for a selection of modes in the superfluid (top row) and supersolid (bottom row) regime. Atoms flow in the direction of the gradient of $f_i$, where red indicates a depletion and blue an increase of atomic density. (b) Energies (in units of the radial trap frequency $\omega_r$) of the BdG excitations as a function of $\epsilon_{dd}$. We label the modes by their circulation quantum number $m$ and show the mode pattern on the superfluid side for the three lowest $m$. We identify a branch of superfluid sound modes (dark blue) and roton modes (purple) on the superfluid side of the transition as well as the first sound branch (green), second sound branch including Goldstone mode (light blue), and Higgs modes (orange) on the supersolid side. A lighter color corresponds to a higher quasimomentum. (c) Speed of sound associated with the first sound as well as the second sound branches of (b) across the transition in units of the sound velocity at the phase transition $c_0 \simeq 2.2\,\mathrm{mm/s}$ (colors as in (b)). The inset shows the superfluid fraction, numerically calculated based on the non-classical rotational inertia.}
	\label{fig:fig1}
\end{figure}

For a direct comparison to the experimental platform of strongly dipolar $^{162}\mathrm{Dy}$ atoms in optical dipole traps, we choose parameters yielding solutions with ${n_d=7}$ droplets \cite{Tengstrand2021}. Scaling relations allow for a  direct transfer of the results to systems with other dipolar strengths, atom numbers, and trapping strengths \cite{Goral2000,Hertkorn2021Spectrum2D,Hertkorn2021Pattern,Schmidt2021}. We assume a torus radius $\rho_0=\SI{1}{\mu m}$, radial trap frequency $\omega_r/2\pi = \SI{1}{kHz}$, aspect ratio $\lambda =1.7$, particle number $N=\num{5e3}$, and dipolar length $a_{dd}=130 a_0$. We calculate the GS for s-wave scatting lengths between $a_s=59a_0$ and $a_s=67a_0$, corresponding to relative dipolar strengths $\epsilon_{dd}=a_{dd}/a_s\in \left[1.95,2.20 \right]$. As a function of $\epsilon_{dd}$ we find the GS to be either superfluid (SF) ($\epsilon_{dd}\lesssim 2.09$) or supersolid (SS). Upon further increasing $\epsilon_{dd}$ the SS eventually develops into a state of $n_d$ isolated droplets. 

As shown in the inset to Fig.~\ref{fig:fig1}(c), we classify the GSs based on the SF fraction. It is unity in the SF and continuously drops in the SS towards higher $\epsilon_{dd}$. The spectrum of the low energy modes across the SF to SS phase transition is shown in Fig.~\ref{fig:fig1}(b), which constitutes one of the main findings of this work. In the SF, each mode is twice degenerate due to the cylindric symmetry \cite{Ronen2006} and the energy of the modes increases monotonically with the circulation number $m$. For a sufficiently large torus radius, where the curvature of the torus is much smaller than the typical length scales of the mode patterns, we can interpret this as a quantization of the one-dimensional momentum scale $q_m=2\pi m/\Theta $, where $\Theta$ is the circumference of toroidal density distribution \footnote{Note that $\Theta \approx 2\pi(1.1\rho_0)$ as the dipolar interaction effectively increases the radius of the SS compared to the actual trap radius \cite{Stringari2023Ring}}. In the long-wavelength limit, we expect a phononic dispersion relation $\omega_m = c_s q_m$, with the SF speed of sound $c_s$. Due to the influence of the dipolar interactions (for the given system size) only the lowest mode ($q=q_1$) falls into this linear regime for the lowest $\epsilon_{dd}$ shown here.
Upon increasing $\epsilon_{dd}$, a number of states with higher $m$ decrease in energy, and can be identified as the angular roton modes \cite{Blakie2012_Roton_a, Blakie2012_Roton_b, Hertkorn2021Spectrum2D,Schmidt2021OblateRoton}. 

At the quantum critical point ($\epsilon^{crit}_{dd}\approx 2.09$) the two $m=7$ modes (degenerate before the transition) touch zero excitation energy and trigger the phase transition from SF to SS, splitting into what we identify as a Higgs and a Goldstone mode \footnote{For brevity, in the following we will refer to the $\tilde{q}=0$ excitations of the SS Higgs and Goldstone branch as Higgs and Goldstone mode, respectively.}. The nature of these modes can be best understood by considering the BdG mode patterns $f_m(x,y)$ displayed in Fig.~\ref{fig:fig1} (a). As both modes emerge from the $m=7$ roton mode, the density fluctuation patterns share the periodicity of the GS. The Goldstone mode pattern is phase-shifted by half a period with respect to the density, while the Higgs pattern is in phase. The two modes correspond to a pure phase and amplitude mode, with striking clarity compared to harmonic 1D systems where phase and amplitude oscillations are always coupled. Due to the periodic boundary conditions, the Higgs and the Goldstone modes are gapless at criticality.

In the SS regime, the GS has developed a periodic density modulation. Accordingly, we can classify the modes by a quasimomentum $\tilde{q}$ folded back into the first Brillouin zone and identify the Higgs and Goldstone modes with $\tilde{q}_0=0$. The SS has three distinct excitation branches: The Higgs branch, with higher-lying $\tilde{q}>0$ modes in addition to the $\tilde{q}_0$ mode, and the first and second sound branches, as indicated in Fig.~\ref{fig:fig1}. The periodic boundary conditions allow us to identify these three branches predicted for infinitely extended SSs \cite{Ferlaino2023Elongated,Ilg2023} in finite and experimentally feasible systems. The Higgs branch consists primarily of amplitude oscillations and rises quickly in energy. The two sound modes show a weaker dependence on $\epsilon_{dd}$, with the first sound branch rising in energy while the second sound branch is decreasing, consistent with a smaller SF fraction \cite{Hofmann2021,Stringari2023Ring}.

Based on the circulation number $m$ and the structure factor (to be discussed below), we can attribute the quasimomentum $\tilde{q}_m=2\pi m/\Theta $, with $m=1,2,3$, to the respective modes. This assignment is supported by employing group theory. The SF GS has the symmetry of the point group $C_{\infty,v}$ describing its invariance under rotation and reflection. At the SF-SS transition this symmetry is broken, $C_{\infty,v}\rightarrow C_{7,v}$ \cite{supmat}, where $C_{7,v}$ contains the irreducible representations $A_1,\ A_2,\ E_1,\  E_2,\  E_3$, 
with symmetries of $A_{1,2}$ corresponding to an $m=7$ angular roton mode, and of $E_j$ to one with $m=j$ compatible with quasimomentum $\tilde{q}_j$.  The allowed quasimomenta, purely based on symmetry arguments, are thus $\tilde{q}=0,1,2,3$. This analysis directly generalizes to an arbitrary droplet number $n_d$ through subduction of $C_{\infty,v}$ to $C_{n_d,v}$ \cite{Altmann1994,Dresselhaus2008,supmat}.

As an example, the mode pattern of the $m=6$ and $m=8$ modes before and after the transition are displayed in Fig.~\ref{fig:fig1} (a). In the SS these modes give rise to the second and first sound at quasimomentum $\tilde{q}_1$ and show a reflection antisymmetry described by $E_1$. We note that in compliance with our observations, the dispersion relation of the two sound modes is expected to be linear for low momenta, before flattening off close to the edge of the Brillouin zone. The Higgs mode energy is a convex function of the quasimomentum. 
Albeit similar excitation spectra were previously reported for finite harmonically trapped systems \cite{Hertkorn2019,Natale2019,Tanzi2019,Chomaz2019,Guo2019}, 
their interpretation in terms of different sound branches was hindered by the absence of well defined quasimomenta and Brillouin zones due to the coupling of linear and COM motion. In contrast, in the toroidal supersolid investigated here, the Higgs modes branch keeps its character even relatively far away from the transition point, as it is not coupled to modes with a different quasimomentum. This allows us to estimate the effective mass of the Higgs excitation \cite{supmat}. While the $\tilde{q}=0$ Higgs mode is exactly zero at the phase transition, even in the few-droplet limit, the finite system size results in a finite Higgs mass at criticality.

Using the energies of the the respective $\tilde{q}_1$ modes we calculate the speed of sound $c_i = \omega_i / \tilde{q}_1 $, where $i=1,2$ corresponds to first and second sound, respectively. The results are shown in Fig.~\ref{fig:fig1} (c) and agree with a recent description in terms of a hydrodynamic model \cite{Hofmann2021, Stringari2023Ring, supmat}. For our parameters, the $m=8$ modes drop below the $m=1$ modes in the SF, causing them to evolve into a first sound and excited Higgs mode in the SS. This reordering is consistent with an avoided crossing of Higgs and first sound branches described in an infinite system at the same density \cite{Ferlaino2023Elongated}. The resulting discontinuity between the SF and first sound at the phase transition point together with the continuous SF fraction, is consistent with the expectation of a second-order phase transition at intermediate densities in the corresponding bulk system. 

\begin{figure}[t!]
	\includegraphics[trim=0 0 0 0,clip,scale=0.45]{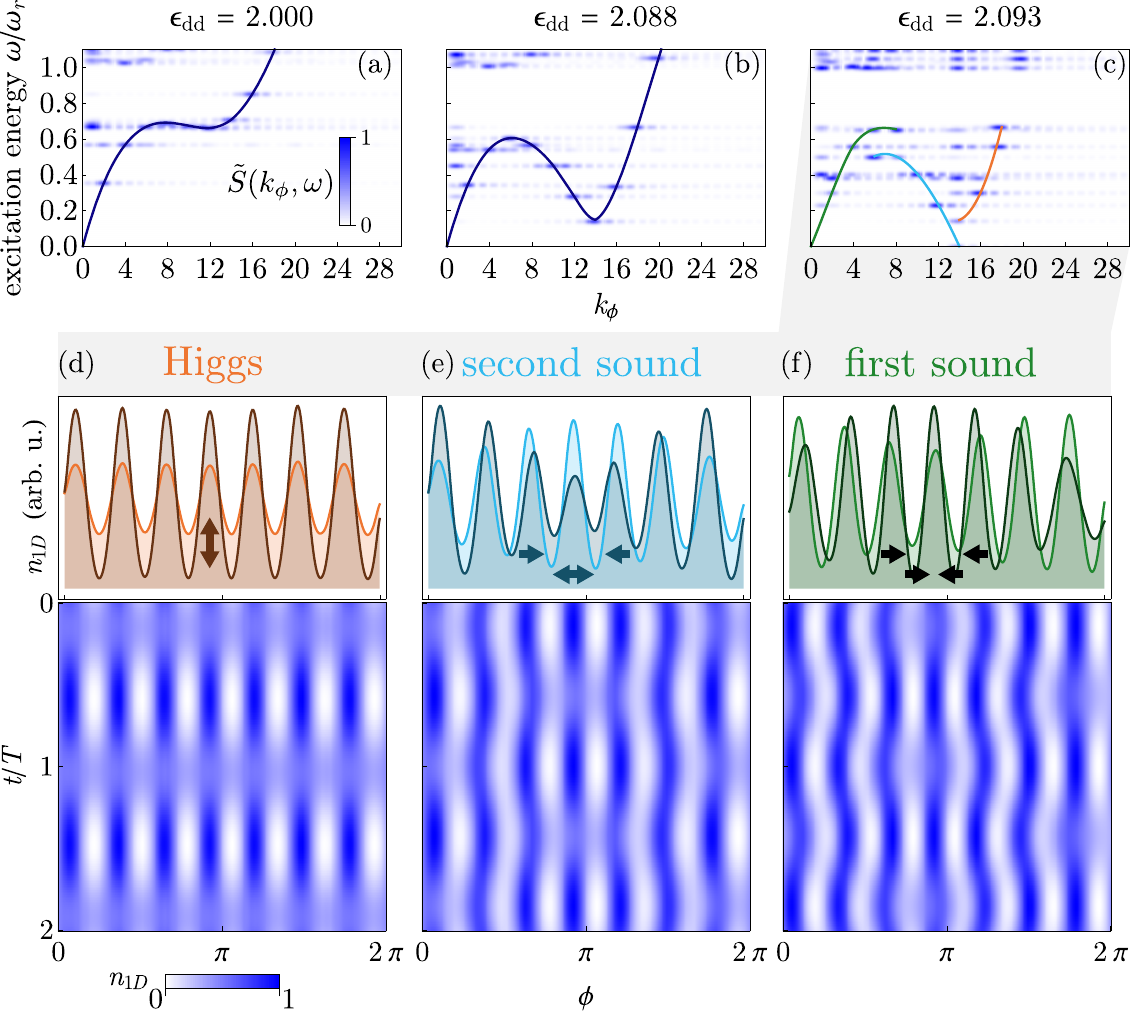}
	\caption{\textbf{Spatial and temporal character of sound and amplitude modes} (a-c) Angular spectrum of the dynamic structure factor ${\tilde{S}(k_\phi,\omega)}$ across the phase transition. We normalize ${\tilde{S}}$ at each $\omega_i$ for visual clarity. Lines through the most prominent peaks of $\tilde{S}$ are guides to the eye. (d-f) The upper panel shows cuts of the density along the torus $n_{1D}(\phi,t)$ at ${t/T = 0,\ 1/2}$ during the time evolution, where ${T=2\pi/\omega}$. The lower panels show the full density evolution $n_{1D}(\phi,t)$ for two periods $T$ of the Higgs mode, second sound, and first sound at $\tilde{q}_1$ from (c). The Higgs mode is an amplitude oscillation mediating a periodic change in the superfluid fraction through superfluid flow into and out of the droplets. The second (first) sound is characterized by an out-of-phase (in-phase) oscillation between the crystal structure and the superfluid background.}
    \label{fig:fig2}
\end{figure}

Our approach allows us to develop a microscopic and intuitive understanding of the character of these elementary excitations (Fig.~\ref{fig:fig2}). To relate our findings to quantities which can be directly measured in experiments \cite{Schmidt2021OblateRoton}, we define the angular spectrum of the structure factor ${ \tilde{S}(k_\phi,\omega) = \int \! \mathrm{d}\phi\, e^{ik_\phi \phi} \tilde{S}(\phi,\omega)}$, shown in Fig.~\ref{fig:fig2}(a)-(c), where $\tilde{S}(\phi,\omega)$ is readily obtained from a Fourier transform of the density fluctuation patterns $\delta n_i = f_i\psi_0$ \cite{Hertkorn2021Fluctuation} in cylinder coordinates at $z=0$ along the torus $r \simeq \rho_0$ \cite{supmat}. A mode with an $m$-fold angular symmetry corresponding to an excitation at momentum $q_m$ shows up as a peak in $\tilde{S}(k_\phi,\omega)$ at ${k_\phi = 2q_m/q_1=2m}$. In the SF regime (Fig.~\ref{fig:fig2}(a)-(b)), $\tilde{S}$ reproduces a discretized version of the known Bogoliubov dispersion relation, which is linear at small $k_\phi$ and quadratic towards larger $k_\phi$. Near the phase transition, a roton minimum develops. On the SS side (Fig.~\ref{fig:fig2}(c)), the SF branch splits up into the first sound, second sound, and Higgs branches.

We find that for the first sound branch, the first mode shows peaks at $k_\phi = 2,\ 14,\ 16$, the second mode at $k_\phi = 4,\ 14,\ 18$, and the third mode at $k_\phi = 6,\ 8,\ 14$. The peak at $k_\phi = 14$ corresponds to a modulation at the spatial frequency of the crystal structure and the other peaks can be mapped to a quasimomentum of $|\tilde{q}_m|$ with $m=1,2,3$. This assignment is consistent with a $m-$fold rotational symmetry of the excitation pattern. In contrast, for the second sound branch, the peak at $k_\phi = 14$ is absent close to the phase transition for all modes. Instead the main peaks are at $12$ ($m=1$), $10$ ($m=2$), and $(6,8)$ for $m=3$. While the character of these modes is strikingly different from the first sound they can be mapped to the same symmetry and quasimomentum. The structure of the Higgs modes similarly maps to a quasimomentum of  $|\tilde{q}_m| = 0,1,2$ \footnote{Apart from the contributions discussed so far, there is a low-frequency contribution at $k_\phi = 1$ which does not correspond to a specific quasimomentum but rather a slow modulation of the envelope of the excitation pattern most likely caused by residual coupling between the two degenerate modes within our numerical framework.}.

It is instructive to study the linearized time evolution  shown for the first mode on each branch, respectively, in Fig.~\ref{fig:fig2}(d)-(f). In Fig.~\ref{fig:fig2}(d), the amplitude character of the Higgs mode becomes apparent. Atoms flow from the SF background into the droplets and vice versa. In contrast, the second (first) sound modes are an out-of-phase (in-phase) oscillation between crystal compression and SF flow (Fig.~\ref{fig:fig2}(e)-(f)). As the crystal compresses towards one node on the torus, the SF density flows towards the opposite (same) direction, reducing (increasing) the density where the droplet spacing is smaller. Due to the dipolar repulsion of the droplets, the compression of the crystal is associated with an energy cost, which is reduced (increased) by the out-of-phase (in-phase) SF flow, providing an intuitive picture of the second sound being slower than the first sound. This zero temperature mode structure is reminiscent of the finite temperature in- and out-of-phase oscillations between SF and normal fluid flow in a (non-dipolar) SF within the two-fluid model \cite{Donnelly2009,Yan2024}. This a-posteriori motivates our interpretation of the corresponding modes in terms of first and second sound.  

\begin{figure}[tb!]
	\includegraphics[trim=0 0 0 0,clip,scale=1]{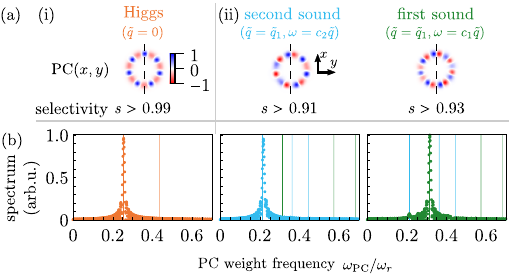}
	\caption{\textbf{Spectroscopy of supersolids} (a) Dominant PCs after applying an excitation scheme (i)-(ii) (see main text) to the GS at $\epsilon_{dd} \simeq 2.1$. (b) Fourier spectra $|\mathcal{F}_w(\omega_\mathrm{PC})|^2$ of the dominant PC weights $w(t)$. Vertical lines in the spectra indicate BdG energies (Fig.~\ref{fig:fig1}(b)). All modes are individually excited with $>90\%$ selectivity. All PCs show above $95\%$ overlap with the density fluctuation pattern obtained from the BdG analysis (Fig.~\ref{fig:fig1}(a)) and oscillate at frequencies corresponding to the BdG energies.}
    \label{fig:fig3}
\end{figure}

Understanding the spatial and temporal character of these elementary excitations as developed in this work helps devise simple and experimentally feasible schemes to selectively excite individual modes (Fig.~\ref{fig:fig3}). We employ principal component analysis (PCA) \cite{Hertkorn2021Fluctuation}, to assess our excitation scheme. 
PCA is a model-free method giving access to the principal components (PCs) and their weights $w(t)$ quantifying their contribution to the spatial variance of the time evolution. We define the selectivity $s$ as the amount of variance described by the time evolution for a given PC. The overlap between the strongest PC and the corresponding BdG mode is above $95\%$ in all cases discussed here.

The Higgs mode (Fig.~\ref{fig:fig3}(i)) can be selectively excited by modulating the s-wave scattering length, which can be done conveniently in state-of-the-art experiments by modulating a magnetic offset in the vicinity of a Feshbach resonance.  This scheme directly couples to the (symmetric) $A_1$ symmetry and allows us to excite the $\tilde{q}=0$ Higgs mode with a selectivity of $>99\%$. 

The Goldstone mode corresponds to a rotation of the system at zero energy in a cylindrically symmetric trap. Any breaking of the cylindrical symmetry elevates this mode to a finite frequency and enables imparting angular momentum into the system. This can be achieved by deforming and rotating the trapping potential or by using magnetostirring \cite{Klaus2022}, where the magnetic field is tilted and rotated. For more details, see \cite{supmat}. 

The two sound modes at $\tilde{q}_1$ (Fig.~\ref{fig:fig3}(ii)) share the symmetry $E_1$. These modes therefore couple to a dipole excitation pattern and we modulate the trap with ${\Delta V \propto x \sin(\omega t)}$ along a symmetry axis $x$ of the SS with the frequency $\omega$ matching the energy of the respective modes. This scheme selectively excites the resonantly driven sound with only a few percent weight in the other sound mode. This selectivity can be further increased by optimizing duration, amplitude, and other details of the modulation scheme. 

Importantly, this scheme cannot couple to the modes at $\tilde{q}_2$ as their symmetry $E_2$ has no overlap with the symmetry $E_1$, while coupling to the $E_3$ mode is possible in principle, but suppressed due to the frequency detuning. Similarly, a quadrupole (hexapole) excitation scheme can be used to selectively excite the $\tilde{q}_2$ ($\tilde{q}_3$) sound modes with no coupling to modes at $\tilde{q}_{1,3}$ (small coupling to modes at $\tilde{q}_{1}$) \cite{supmat}. Based on the symmetry considerations, one can generalize these observations to a set of selection rules, establishing a complete toolbox for the spectroscopy of supersolids. The absence of linear and COM motion coupling ensures that a mode retains its character over time and a range of interaction strengths and is the basis of such spectroscopy.

In conclusion, we have uncovered the origin of second sound and Higgs amplitude modes in dipolar supersolids by employing a toroidal trap, establishing a correspondence between elementary excitations of finite and bulk supersolid systems. Strikingly, toroidal supersolids in the low-momentum limit realize an isolated and massive Higgs excitation that should propagate as a massive quasiparticle in the self-generated periodic structure of the supersolid. Scattering of the Higgs excitation at an obstacle, such as a weak link on the torus, and its dispersion in larger tori is the starting point for studies of massive quasiparticles in supersolids. Our work highlights the analogy between the second sound in supersolids at zero temperature and second sound in superfluids at finite temperature, which are out-of-phase oscillations between superfluid and non-superfluid components. We believe that future studies considering finite temperature supersolids \cite{Aybar2019,Sanchez-Baena2023} in toroidal traps could provide insight into the origin of further sound modes arising through the presence of two non-superfluid (crystal and thermal) components.

\section*{Acknowledgments}
We acknowledge discussions and input in particular from T. Langen, as well as M. Nilsson Tengstrand at an earlier stage of the project, and also thank T. Arnone Cardinale, L. Chergui and H. P. Büchler for discussions. 
P.S., K.M. and S.M.R. gratefully acknowledge financial support from  the Knut and Alice Wallenberg Foundation (KAW 2018.0217) and the Swedish Research Council (grant no. 2022-03654). J.H., K.S.H.N., P.U., F.H., L.L., T.P. and R.K. acknowledge support from the European Research Council (ERC) (grant agreement No. 101019739). J.H. gratefully acknowledges financial support by the Vector Stiftung (project no. P2021-0114). L.L. acknowledges funding from the Alexander von Humboldt Foundation. Part of the computations were enabled by resources provided by the National Academic Infrastructure for Supercomputing in Sweden (NAISS), partially funded by the Swedish Research Council through grant agreement no. 2022-06725.

\bibliography{biblio}

\newpage
\section*{Supplementary Material}
\subsection{Details on the numerically calculated excitation spectrum}
We calculate the GS of the dipolar toroidal system using the eGPE framework previously described e.\,g. in \cite{Wenzel2017,Hertkorn2019} and gain access to the excitation spectrum using the BdG framework described in more detail in e.\,g. \cite{Guo2019,Hertkorn2019,Hertkorn2021Fluctuation}.
We have chosen the parameters used in this study to ensure that the $n_d=7$ droplet solution is the ground state for any $\epsilon_{dd}$ considered. However, for larger $\epsilon_{dd}\gtrsim 2.15$, the $n_d=8$ solutions become very similar in energy, such that numerically a bistability between $n_d=7$ and $n_d=8$ solutions develops. In this work, we restrict ourselves to $n_d=7$. In the spectrum, we focus on the low-energy behavior of longitudinal excitations along the torus ($\omega < \omega_r$). The Goldstone mode numerically shows a finite energy $\omega/\omega_r <0.01$ for the spatial grid size employed here ($96^3$ grid points). We have confirmed that this energy decreases for finer spatial grids and tighter convergence criteria, consistent with a gapless Goldstone mode.\\
We also note that for the torus radius of $\rho=\SI{1}{\mu m}$ used in this work, there is still a small residual dipolar coupling across the torus. We have however explicitly checked by calculating spectra with a scaled torus size and atom number, that such scaling does not qualitatively change any of the results presented in this work.

\subsection{Superfluid speed of sound and compressibility}
In a hydrodynamic description \cite{PitaevskiiBook2016} the speed of sound 
\begin{equation}
c_\kappa = \frac{1}{\sqrt{M\bar{n}\kappa}}
\end{equation}
is defined based on the compressibility
\begin{equation}
\kappa = \frac{1}{\bar{n}^2}\frac{\partial \bar{n}}{\partial \mu}
\end{equation}
where $\mu$ is the chemical potential and $\bar{n}$ is the average density. The compressibility is a GS property, providing a definition of this speed of sound independent of the BdG analysis. In the SF, $c_\kappa$ coincides with the SF speed of sound $c_s$ \cite{PitaevskiiBook2016}. 

For comparison to our BdG analysis we calculate $c_\kappa$ in the SF. To this end we note that $\bar{n}\kappa$ is independent of the unit of $\bar{n}$, and so is $c_\kappa$. Therefore we can write ${\bar{n}\kappa = \mathrm{d}\ln \bar{n} /\mathrm{d}\mu}$. The average azimuthal density for a torus with circumference $\Theta$ is defined by $\bar{n}=N/\Theta$ which is independent of the detailed spatial dependence of the density $n(x,y,z)$. We find that $\Theta/2\pi$ varies by less than $0.01\,\mathrm{\mu m}$ in the SF for our parameters, allowing to a good approximation to calculate $c_\kappa$ using ${\bar{n}\kappa \approx \mathrm{d}\ln N/\mathrm{d}\mu}$. To estimate $\bar{n}\kappa$ we vary $N$ by a few percent, calculate the GS, and obtain $\mu(\bar{n})$. Using either method we obtain reasonable agreement with the BdG description, with $c_\kappa$ typically $\simeq 20\%$ larger compared to the BdG speed of sound. Similar deviations have been reported \cite{Stringari2023Ring} and are likely due to the finite torus size.
Very recent developments on the hydrodynamic description of supersolids also provide expressions for the speeds of sound in the SS \cite{Dorsey2010,Hofmann2021,Stringari2023Ring,Platt2024} depending on $c_\kappa$, the SF fraction $f_s$, and the layer compressibility modulus $B$.

\subsection{Higgs effective mass}
\begin{figure}[tb!]
	\includegraphics[trim=0 0 0 0,clip,scale=1.0]{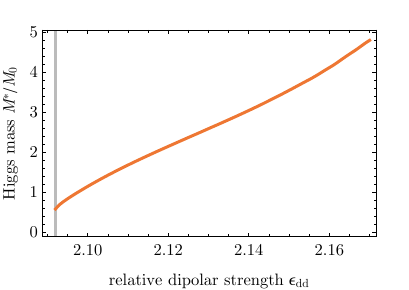}
	\caption{\textbf{Effective mass of the Higgs mode.} The effective mass of the amplitude mode, calculated based on the lowest two energy levels of the corresponding Higgs branch, is plotted as a function of $\epsilon_{dd}$. We observe a finite mass even at criticality and an increase in the effective mass, when the superfluid fraction decreases towards the regime of isolated droplets. The vertical gray line indicates the phase transition point.}
	\label{fig:figHiggs}
\end{figure}
In previous studies on 1D supersolids in harmonic traps, where the linear quasimomentum couples to the center of mass momentum, the Higgs character of the amplitude mode was maintained only very close to the phase transition \cite{Hertkorn2019}. In stark contrast, in the toroidal geometry studied here, we observe the Higgs modes at $\tilde{q}_m=2\pi m / \Theta$ with $m=1,2,3$ to be isolated over the whole $\epsilon_{dd}$ regime studied in this work. This isolation allows us to define an effective mass $M^*$ of the excitation. The effective mass describes the dispersion of a localized Higgs quasiparticle wavepacket excited at some position on the torus. To estimate the mass we define the energy difference between the $\tilde{q}_0$ and $\tilde{q}_1$ Higgs mode as $\hbar \Delta \omega_{H}$ and calculate the effective mass via $M^*=\hbar^2 \tilde{q}_1^2/2 \hbar \Delta \omega_{H}$. Here, we only use the lowest two Higgs energy levels as the $\tilde{q}_2$ mode is already close to the edge of the Brillouin zone. The Higgs effective mass is shown as a function of $\epsilon_{dd}$ in Fig~\ref{fig:figHiggs}. In accordance to its behaviour in bulk systems, we observe the effective mass to increase when reducing the SF fraction. In the limit of large system sizes, we expect the effective mass to approach zero at the phase transition point. Due to the finite size of the system and correspondingly the finite sampling of allowed quasimomenta, we observe a\\\\

\onecolumngrid

\begin{table}[h]
\caption{Compatible irreducible representations between the angular roton modes of the superfluid to elementary excitations of the supersolid with $n_d$ droplets when the rotational symmetry is spontaneously broken $C_{\infty,v} \rightarrow C_{n_d,v}$. A section of the full table which continues indefinitely to the right and bottom is shown. The row for the symmetry $C_{7,v}$ of the ground state in the main text is highlighted. The pattern emerging in the body of the table towards higher $n_d$ along the rows continues predictably and has a physical interpretation, see main text.}
	\label{tab:compat}
	\centering
	\resizebox{\textwidth}{!}{%
		\begin{tabular}{||c| c| c| c| c| c| c| c| c| c| c| c| c| c| c||} 
			\hline
			$m$ ($C_{\infty,v}$) & 0 & 1 & 2 & 3 & 4 & 5 & 6 & 7 & 8 & 9 & 10 & 11 & 12 & 13\\
			\hline\hline
			$C_{1,v}$ & $A_1$ & $A_1+A_2$ & $A_1+A_2$ & $A_1+A_2$ & $A_1+A_2$ & $A_1+A_2$ & $A_1+A_2$ & $A_1+A_2$ & $A_1+A_2$ & $A_1+A_2$ & $A_1+A_2$ & $A_1+A_2$ & $A_1+A_2$ & $A_1+A_2$\\
			\hline
			$C_{2,v}$ & $A_1$ & $B_1+B_2$ & $A_1+A_2$ & $B_1+B_2$ & $A_1+A_2$ & $B_1+B_2$ & $A_1+A_2$ & $B_1+B_2$ & $A_1+A_2$ & $B_1+B_2$ & $A_1+A_2$ & $B_1+B_2$ & $A_1+A_2$ & $B_1+B_2$ \\
			\hline
			$C_{3,v}$ & $A_1$ & $E$ & $E$ & $A_1+A_2$ & $E$ & $E$ & $A_1+A_2$ & $E$ & $E$ & $A_1+A_2$ & $E$ & $E$ & $A_1+A_2$ & $E$ \\
			\hline
			$C_{4,v}$ & $A_1$ & $E$ & $B_1+B_2$ & $E$ & $A_1+A_2$ & $E$ & $B_1+B_2$ & $E$ & $A_1+A_2$ & $E$ & $B_1+B_2$ & $E$ & $A_1+A_2$ & $E$ \\
			\hline
			$C_{5,v}$ & $A_1$ & $E_1$ & $E_2$ & $E_2$ & $E_1$ & $A_1+A_2$ & $E_1$ & $E_2$ & $E_2$ & $E_1$ & $A_1+A_2$ & $E_1$ & $E_2$ & $E_2$ \\
			\hline
			$C_{6,v}$ & $A_1$ & $E_1$ & $E_2$ & $B_1+B_2$ & $E_2$ & $E_1$ & $A_1+A_2$ & $E_1$ & $E_2$ & $B_1+B_2$ & $E_2$ & $E_1$ & $A_1+A_2$ & $E_1$ \\
			\hline
			$\boldsymbol{C_{7,v}}$ & $\boldsymbol{A_1}$ & $\boldsymbol{E_1}$ & $\boldsymbol{E_2}$ & $\boldsymbol{E_3}$ & $\boldsymbol{E_3}$ & $\boldsymbol{E_2}$ & $\boldsymbol{E_1}$ & $\boldsymbol{A_1+A_2}$ & $\boldsymbol{E_1}$ & $\boldsymbol{E_2}$ & $\boldsymbol{E_3}$ & $\boldsymbol{E_3}$ & $\boldsymbol{E_2}$ & $\boldsymbol{E_1}$ \\
			\hline
			$C_{8,v}$ & $A_1$ & $E_1$ & $E_2$ & $E_3$ & $B_1+B_2$ & $E_3$ & $E_2$ & $E_1$ & $A_1+A_2$ & $E_1$ & $E_2$ & $E_3$ & $B_1+B_2$ & $E_3$\\
			\hline
			$C_{9,v}$ & $A_1$ & $E_1$ & $E_2$ & $E_3$ & $E_4$ & $E_4$ & $E_3$ & $E_2$ & $E_1$ & $A_1+A_2$ & $E_1$ & $E_2$ & $E_3$ & $E_4$ \\
			\hline
			$C_{10,v}$ & $A_1$ & $E_1$ & $E_2$ & $E_3$ & $E_4$ & $B_1+B_2$ & $E_4$ & $E_3$ & $E_2$ & $E_1$ & $A_1+A_2$ & $E_1$ & $E_2$ & $E_3$ \\
			\hline
			$C_{11,v}$ & $A_1$ & $E_1$ & $E_2$ & $E_3$ & $E_4$ & $E_5$ & $E_5$ & $E_4$ & $E_3$ & $E_2$ & $E_1$ & $A_1+A_2$ & $E_1$ & $E_2$ \\
			\hline
			$C_{12,v}$ & $A_1$ & $E_1$ & $E_2$ & $E_3$ & $E_4$ & $E_5$ & $B_1+B_2$ & $E_5$ & $E_4$ & $E_3$ & $E_2$ & $E_1$ & $A_1+A_2$ & $E_1$ \\
			\hline
			$C_{13,v}$ & $A_1$ & $E_1$ & $E_2$ & $E_3$ & $E_4$ & $E_5$ & $E_6$ & $E_6$ & $E_5$ & $E_4$ & $E_3$ & $E_2$ & $E_1$ & $A_1+A_2$ \\
			\hline
		\end{tabular}
	}

\end{table}

\twocolumngrid

small finite Higgs mass given by ${M^*\simeq 0.104 M(^{162} \mathrm{Dy}) \simeq 0.58 M_0}$ even at the phase transition, where ${M_0 = \hbar \tilde{q}_1/c_0 \simeq 29\,\mathrm{u}}$. Here, $M_0$ is defined via the SF speed of sound $c_0$ at the critical point resulting in the convenient dimensionless expression $M^*/M_0=\omega_0/(2\Delta \omega_H)$ with $\omega_0 = c_0 \tilde{q}_1$.

\subsection{Superfluid fraction}
The spontaneous breaking of the continuous rotational symmetry, in addition to the already broken $U(1)$ symmetry, results in periodic density modulations and a supersolid phase. A common way to measure the interplay between solid and SF 'components' is via the SF fraction $f_s$. While frequently calculated via an upper and and lower bound $f_s^\pm$ introduced by Legget \cite{Leggett1998}, our annular geometry permits calculation of the SF fraction via the system's moment-of-inertia and response to external rotation as \cite{Leggett1970,Tengstrand2021,Tengstrand2023,Stringari2023Ring}
\begin{align}
f_s = 1-\lim_{\Omega \to 0} \frac{\langle\hat{L}_z \rangle}{N M\langle\rho^2\rangle\Omega},
\end{align}
where $\langle\rho^2\rangle$ is evaluated numerically. $\Omega$ denotes the angular frequency of rotation (herein, $\Omega=10^{-7}\omega_r$). We have explicitly checked, that the SF fraction obtained this way (shown in Fig.~\ref{fig:fig1} of the main text) is consistent with the $f_s^\pm$ Legget bounds.

\subsection{Group theory} \label{sec:gt}

Here, we provide some additional details on the group $C_{n_d, v}$, describing the modes of a supersolid with $n_d$ droplets in a cylindrically symmetric trap. Our discussion concerning groups closely follows standard group theory texts \cite{Dresselhaus2008}, to which the interested reader is referred for further information. Our description extends and generalizes the understanding obtained for one-dimensional and two-dimensional supersolids in harmonic traps with small droplet numbers \cite{Guo2019,Hertkorn2019,Hertkorn2021Spectrum2D}.\\
The $m=n_d$ angular roton modes drive the SF to SS phase transition breaking the continuous rotational symmetry $C_{\infty,v} \rightarrow C_{n_d,v}$. In Tab.~\eqref{tab:compat} we tabulate which $m-$angular roton modes are compatible with which irreducible representation across every point group (up to $m=n_d=13$). We obtain this table by considering that $E_i$ transform as $m=i$ angular roton modes. The periodicity in the pattern in Tab.~\ref{tab:compat} has a physical interpretation. When the symmetry is broken $C_{\infty,v} \rightarrow C_{n_d,v}$ momenta that were continuously increasing corresponding to the angular roton modes with $m$ up to $\infty$ can be folded back to the first Brillouin zone in which only momenta up to $m=n_d/2$ for even $n_d$ and $m=(n_d-1)/2$ for odd $n_d$ are available. Therefore angular roton modes are assigned to $E_i$ with increasing $i$ for $i<m/2$, to $B_1 + B_2$ right at the edge of the Brillouin zone $i=m/2$, and to $E_i$ with decreasing $i$ for $i>m/2$. For example in $C_{4,v}$ modes compatible with $E$ transform as $m=1$ angular roton modes and modes compatible with $B_1,B_2$ transform as $m=2$ angular roton modes. Angular roton modes with $m=1,3,5,...$ remain degenerate ($E$) and those with $m=2,6,10,...$ split up into two levels ($B_1,B_2$).
To discuss the symmetries in a specific group  we exemplify $n_d=7$ case. The symmetries of the resulting supersolid GS can be described by the point group $C_{7,v}$. The character table of the group $C_{7,v}$ is reproduced in Tab.~\eqref{tab:C7v}, where classes are arranged along the columns, irreducible representations along the rows and the body of the table shows their character. The supersolid is symmetric with respect to the identity ($E$), rotation ($C_7$) around $\pm 2\pi/7$, rotation ($C_7^2$) around $\pm 4\pi/7$, rotation ($C_7^3$) around $\pm 6\pi/7$ around the $z$-axis, and reflection $(\sigma_v)$ through the seven vertical planes intersecting the $z$-axis and the droplets. These five types of symmetries divide the group $C_{7,v}$ into the five classes $E$, $2C_7$, $2C_7^2$, $2C_7^3$, $7\sigma_v$ and yield by the orthogonality theorem five irreducible representations $A_1$, $A_2$, $E_1$, $E_2$, and $E_3$. $A_i$ and $E_i$ are one-dimensional and two-dimensional irreducible representaitons, respectively. The notations used for the symmetries and irreducible representations are known as Schoenflies and Mulliken symbols, respectively. Defining the character as the trace of the matrix of a representation and using that the character for each element in a class is the same, essential information of a group can be summarized in a character table. The statements can be easily adapted for any point group $C_{n_d,v}$ once the character table of the point group is known. The character tables for many point groups are tabulated \cite{Altmann1994,Dresselhaus2008}.

\bgroup
\def\arraystretch{1.3} % row width
\begin{table}[htb!]
	\caption{Character table for the group $C_{7,v}$. The irrotational character values are ${\chi_n = 2 \cos(2\pi n/7)}$.}
	\label{tab:C7v}
	\begin{center}
		\begin{tabular}{ >{\centering}p{0.6cm} || >{\centering\arraybackslash}p{0.5cm} | >{\centering\arraybackslash}p{0.5cm} | >{\centering\arraybackslash}p{0.5cm} | >{\centering\arraybackslash}p{0.5cm} | >{\centering\arraybackslash}p{0.5cm} || >{\centering\arraybackslash}p{4cm} }
			$C_{7,v}$ & $E$ & $2C_7$& $2C_7^2$ & $2C_7^3$ & $7\sigma_v$ & example basis\\
			\hline\hline
			$A_1$ & 1 & 1 & 1 & 1 & 1 &$z$\\
			$A_2$ & 1 & 1 & 1 & 1 & -1 &$R_z$\\
			$E_1$ & 2 & $\chi_1$ & $\chi_2$ & $\chi_3$ & 0       &$\{x, y \}$\\
			$E_2$ & 2 & $\chi_2$ & $\chi_3$ & $\chi_1$ & 0 &$\{x^2-y^2, xy\}$\\
			$E_3$ & 2 & $\chi_3$ & $\chi_1$ & $\chi_2$ & 0 & $\{x(x^2-3y^2), y(3x^2-y^2) \}$
		\end{tabular}
	\end{center}	
\end{table}
\egroup

To get an intuition on the symmetries compatible with each of the irreducible representations, we consider the example basis functions given in Tab.~\ref{tab:C7v}. Modes compatible with $A_1$ transform as $z$ -- symmetric with respect to all rotations and reflections in the group. Modes compatible under $A_2$ transform as $R_z$ (rotation around the $z$-axis) -- antisymmetric with respect to $7\sigma_v$ and symmetric otherwise. The $m=7$ angular roton modes therefore split into a zero energy Goldstone mode compatible with $A_2$ and a Higgs amplitude mode compatible with $A_1$. The Goldstone mode corresponds to the $q=0$ limit of the second sound branch and is a rotation of the system. An alternative point of view is that this mode is not physical because it lies at $\omega = q = 0$. We find that the identification of the zero energy Goldstone mode as the first instance on the second sound branch is nonetheless useful, in particular since this interpretation is consistent with the lowest energy Goldstone mode being second sound in harmonic traps, lifted to a finite energy due to the already broken translational symmetry. The Higgs mode is an amplitude modulation of the supersolid order parameter, where the SF and crystal part exchange density in phase. The two-dimensional representations $E_i$ share the same symmetry with the $m=i$ angular roton modes. For example, modes compatible with $E_1$ are either antisymmetric with respect to $x$ or $y$, exactly like the two degenerate $m=1$ angular roton modes. This identification allows generally to think of the $m=i$ angular roton modes as basis functions, representing the symmetry of modes compatible with $E_i$. $E$ in $C_{3,v}$ and $C_{4,v}$ transforms just as $E_1$ in the point groups of higher symmetry.

Finally, we comment on the correspondence to the one-dimensional droplet chains in harmonic traps \cite{Guo2019,Hertkorn2019}. The group $C_{1,v}$ is identical to $C_s$ which is isomorphic to $C_2$ and the symmetric group $S_2$. Modes can therefore be classified according to their reflection symmetry around the center of the trap and will be either symmetric and compatible to $A_1$ or antisymmetric and compatible to $A_2$. Whether specific modes correspond to first or second sound can be identified by the eigenenergies increasing (first sound) or decreasing (second sound) towards larger $\epsilon_{dd}$. Avoided crossings are expected whenever modes of the same symmetry are close in energy. This holds in particular in systems with low symmetry such as small droplet numbers in one-dimensional and two-dimensional harmonic traps, which explains previous observations showing many avoided crossings \cite{Hertkorn2019,Guo2019,Hertkorn2021Spectrum2D}.

\subsection{Structure factor}
The dynamic structure factor \cite{Hertkorn2021Fluctuation} is given by
\begin{equation}
S(\boldsymbol{k}, \omega) = \sum_i |\delta n_i(
\boldsymbol{k})|^2 \delta (\omega-\omega_i)
\end{equation}
where $\delta n_i(\boldsymbol{k})$ is the Fourier transform of the density fluctuation pattern $\delta n_i(\boldsymbol{r}) = f_i^*(\boldsymbol{r})\psi_0(\boldsymbol{r})$ with the Bogoliubov mode pattern $f_i = u_i + v_i$ and the GS wavefunction $\psi_0$. As we focus on angular excitations and thus the $x-y$ plane, we transform the structure factor to polar coordinates ${S(x,y,\omega) \rightarrow S(k, \phi,\omega)}$. We then fix $k$ to the radial momentum at which $S$ displays a maximum and are left with the angular distribution of the structure factor $\tilde{S}(\phi,\omega) = S(\phi,\omega)-\overline{S}$ where $\overline{S}$ indicates the average of $S(\phi,\omega)$ along $\phi$. An angular roton mode with $m$ nodes will result in $2m$ maxima or equivalently $2m$ full modulations in $\tilde{S}(\phi,\omega)$. To assign a quasimomentum to these modes and classify the more complicated shapes of $f_i$ in the supersolid phase, we Fourier-transform $\tilde{S}(\phi,\omega)$ along the angular direction
\begin{equation}
\tilde{S}(k_\phi,\omega) = \int \! \mathrm{d}\phi\, e^{ik_\phi \phi} \tilde{S}(\phi,\omega)
\end{equation}
where $k_\phi$ is the angular frequency. In this way, $2m$ maxima along $S(\phi,\omega)$ for $\phi$ from $0$ to $2\pi$ correspond to an angular frequency of $k_\phi = 2m$. In conclusion, a peak in $\tilde{S}(k_\phi,\omega)$ at $k_\phi = 2m$ corresponds to $2m$ angular modulations in the structure factor $S(\boldsymbol{k}, \omega)$ and indicates an excitation with an angular symmetry compatible with $m$ modulations, such as a $m$-angular roton mode or a mode compatible with $E_m$ in the group $C_{n_d,v}$, see Sec.~\ref{sec:gt}.

In Fig.~1 of the main text, we broaden $\tilde{S}(k_\phi,\omega)$ along $\omega$ and normalize at each $\omega_i$ by the maximum of $\tilde{S}$ along $k_\phi$ for visual clarity. We use $\tilde{S}$ to assign quasimomenta, follow the dispersion relation across the phase transition, and relate the spectrum to the dispersion relation of infinitely extended supersolids. The amplitude of $\tilde{S}$ plays a role in evaluating sum rules based on $S(\boldsymbol{k}, \omega)$ \cite{PitaevskiiBook2016,Hertkorn2021Fluctuation,Stringari2023Ring}, but is not considered any further here.

\subsection{Spectroscopy of supersolids}
\begin{figure}[t!]
	\includegraphics[trim=0 0 0 0,clip,scale=1.0]{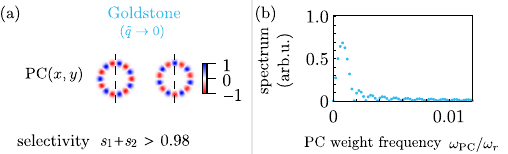}
	\caption{\textbf{Exciting the Goldstone mode}. (a) Dominant PCs after magnetostirring the GS at $\epsilon_{dd}\simeq 2.1$. (b) Fourier spectrum $|\mathcal{F}_w(\omega_\mathrm{PC})|^2$ of the dominant PC weights $w(t)$. }
	\label{fig:Goldstone}
\end{figure}

The toroidal trap protects the individual modes from coupling to each other, allowing spectroscopic methods to selectively excite individual modes of the supersolids. Here, we provide additional details on the dynamic protocols used in the main text. Starting with the ground state at $\epsilon_{dd} \simeq 2.1$, we apply a modulation scheme for a certain period of time and then time-evolve the state for typically $100\,\mathrm{ms}$ (or longer for the low-energy Goldstone mode).

To excite the Higgs amplitude mode at $\tilde{q}_0$ we modulate the scattering length ${a_s(t)=a_{s,mod}\sin(\omega_{mod}t)}$  with an amplitude of ${a_{s,mod} =0.05\,a_0}$ for a duration ${t_{mod}=4T_{mod}}$ with $T_{mod}=2\pi/\omega_{mod}$ and the driving frequency $\omega_{mod}/2\pi = 50\,\mathrm{Hz}$. We have chosen a frequency that does not match the Higgs frequency, highlighting that the excitation of the $\tilde{q}_0$ mode does not require resonant driving. With further simulations using this modulation scheme with different parameters, we confirmed that the similar selectivity and spectral purity of the Higgs mode is reached with frequencies below $\omega_{mod}/2\pi = 100\,\mathrm{Hz}$. Eventually at sufficiently high modulation frequencies ($\gtrsim 200 \, \mathrm{Hz}$), further modes are excited as well. In these cases, the PCs become superpositions of BdG modes and the corresponding (multiple) BdG frequencies show up in the spectra of the PC weights. A similar effect is observed towards higher amplitudes $a_{s,mod} \gg 0.1\,a_0$.

The Goldstone mode (Fig.~\ref{fig:Goldstone}) corresponds to a rotation of the system at zero energy in a cylindrically symmetric trap. Breaking the cylindrical symmetry elevates this mode  to a finite frequency and enables imparting angular momentum into the system. This can be done for example by deforming and rotating the trapping potential or by using magnetostirring \cite{Klaus2022} where the magnetic offset field is tilted and rotated. Using the latter, PCA captures the resulting slow rotation by two PCs, identical up to rotation of $2\pi/14$, oscillating in time $\pi/2$ out of phase. Each PC matches the Goldstone mode pattern and the symmetry $A_2$. In order to further validate this approach, we have explicitly calculated the excitation spectrum, for a tilt of the magnetic field axis of $10^\circ$ with respect to the axial direction. This is the tilt employed in the magnetostirring in Fig.~\ref{fig:Goldstone}. For these small tilt angles, the Goldstone mode is still well isolated for all values of $\epsilon_{dd}$ and the overall structure of the spectrum remains intact. The main qualitative difference is that twofold degenerate modes split up as the rotational symmetry is broken with a tilted magnetic field.

The first and second sound modes at $\tilde{q}_1$ are excited by modulating a linear gradient ${\Delta V = p x \sin(\omega t)}$ on top of the toroidal trap where $x$ corresponds to the direction of a symmetry axis of the SS. The perturbation $x$ is chosen since it shares the same symmetry with the basis functions of $E_1$. We choose an amplitude of ${p\simeq 6.3 \times 10^{-3} \omega_r/\rho_0}$ ${(p/2\pi\hbar = 6.3\,\mathrm{Hz}/\mathrm{\mu m})}$ and modulate at the resonance frequency $\omega_i = c_i \tilde{q}_1$ for a time $t_{mod} = 4T_{mod}$ for first sound $(i=1,\ \omega_1/2\pi \simeq 212\,\mathrm{Hz})$ and second sound $(i=2,\ \omega_2/2\pi \simeq 315\,\mathrm{Hz})$. The scheme is robust with respect to the specific values chosen for $t_{mod}$, $\omega$, and $p$. Deviations of $\omega$ from $\omega_i$, of $x$ from the symmetry axis, and towards large $p$ result increasingly in the excitation of superpositions of first and second sound. With a misalignment of $x$ from the symmetry axis it is also possible to set the SS into rotation (excitation of the Goldstone mode). We find similar behavior, although slightly worse selectivity ($60$-$90\%$ depending on modulation parameter details), when modulating a quadrupole field ${\Delta V \propto xy\sin(\omega t)}$ on top of the toroidal trap in order to excite the sound modes at $\tilde{q}_2$.

\end{document}